\documentclass[twocolumn,showpacs]{revtex4}
\usepackage{amsfonts}
\usepackage{amsthm}
\usepackage{amssymb}
\usepackage{amscd}
\usepackage{graphicx}
\input{psfig.sty}

\newcommand{\bastar}{\begin{eqnarray*}}
\newcommand{\eastar}{\end{eqnarray*}}
\newskip\humongous \humongous=0pt plus 1000pt minus 1000pt

\newif\ifdtup

\relax
\newcommand{\be}{\begin{equation}}
\newcommand{\ee}{\end{equation}}
\newcommand{\bea}{\begin{eqnarray}}
\newcommand{\eea}{\end{eqnarray}}
\newcommand{\X}{{\vec X}}
\newcommand{\pro}{\partial}
\newcommand{\n}{\hat n}
\newcommand{\oneg}{\displaystyle\frac{1}{g}}

\newcommand{\D}{{\hat D}}

\newcommand{\A}{{\vec A}}
\newcommand{\valpha}{{\vec \alpha}}

\newcommand{\hn}{{\hat n}}
\newcommand{\hD}{{\hat D}}
\newcommand{\dfrac}{\displaystyle\frac}
\newcommand{\ba}{\begin{array}}
\newcommand{\ea}{\end{array}}

\newcommand{\nn}{\nonumber}
\newcommand{\tD}{\tilde \Delta}
\begin{document}
\title{Monopole Condensation and Dimensional Transmutation in SU(2) QCD}
\bigskip

\author{Y. M. Cho}
\email{ymcho@yongmin.snu.ac.kr}
\author{M. L. Walker}
\email{mlwalker@phya.snu.ac.kr}
\affiliation{Department of Physics, College of Natural Sciences,
Seoul National University, Seoul 151-747, Korea }
\author{D. G. Pak}
\email{dgpak@mail.tps.uz}
\affiliation{Institute of Applied Physics, National University
of Uzbekistan,\\ Tashkent 700-174, Uzbekistan}
\date{}
\begin{abstract}
We resolve the controversy on the stability of the monopole condensation
in the one-loop effective action of $SU(2)$ QCD by calculating
the imaginary part of the effective action
with two different methods at one-loop
order. Our result confirms that the effective action for
the magnetic background has no imaginary part but the one for
the electric background has a negative imaginary part. This
assures that the monopole condensation is indeed stable,
but the electric background becomes unstable due to the pair-annihilation
of gluons.
\end{abstract}
\maketitle

\section{Introduction}

One of the most outstanding problems
in theoretical physics is the confinement problem in
quantum chromodynamics (QCD). It has
long been argued that monopole condensation can explain the
confinement of color through the dual Meissner effect
\cite{nambu,cho1}.
Indeed, if one assumes monopole condensation, one can easily argue
that the ensuing dual Meissner effect guarantees the confinement
\cite{cho1,cho2}. A satisfactory theoretical proof of the desired
monopole condensation in QCD, however, has been very 
elusive \cite{savv,niel,ditt}. Fortunately, we have recently been able 
to demonstrate the monopole condensation with 
the one-loop effective action in $SU(2)$ QCD \cite{cho3,cho4}. 
In this paper we discuss the perturbative method which
can prove the stability 
of the monopole condensation in detail. Our result strongly
indicates that the monopole condensation describes 
the stable vacuum of QCD, which could guarantee the confinement
of color. 

A natural way to establish the monopole condensation in QCD
is to demonstrate that the quantum fluctuation triggers
a dimensional transmutation and phase transition through
the Coleman-Weinberg mechanism in QCD. 
Coleman and Weinberg have demonstrated that
the quantum effect could trigger a dimensional transmutation
in massless scalar QED with a quartic self-interaction of
scalar electrons, by showing that the one-loop effective action generates
a mass gap through the condensation of scalar electrons which defines 
a non-trivial new vacuum \cite{cole}.
To prove the monopole condensation, one need to demonstrate
such a phase transition in QCD.
There have been many attempts to demonstrate monopole condensation
in QCD with the one-loop effective action using the background
field method \cite{savv,niel,ditt}. Savvidy has first calculated 
the effective action of $SU(2)$ QCD in the presence of an {\it ad hoc}
color magnetic background, and discovered an encouraging 
evidence of a magnetic condensation as a non-trivial vacuum
of QCD \cite{savv}. But this calculation was repeated by Nielsen and 
Olesen, who found that the magnetic background 
generates an imaginary part to the effective action
which makes the vacuum unstable \cite{niel}.  
The origin of this instability of the ``Savvidy-Nielsen-Olesen (SNO)
vacuum'' can be traced back to
the fact that it is not gauge invariant.
Because of this the effective action develops an imaginary part
which destabilizes the magnetic condensation through
the pair creation of gluons. This instability of the
SNO vacuum has been widely accepted and never been seriously
challenged. This has created the unfortunate impression that
it might be impossible to establish the monopole condensation
with the effective action of QCD.

A few years ago, however, there was a new attempt to calculate
the one-loop effective action of QCD
with a gauge independent separation
of the classical background from the quantum field \cite{cho3}.
Remarkably, in this
calculation the effective action was shown to
produce no imaginary part in the presence of
the non-Abelian monopole background, but a negative imaginary part
in the presence
of the pure color electric background. This means that in QCD
the non-Abelian monopole background produces a stable monopole
condensation, but the color electric background
becomes unstable by generating a pair annhilation of
the valence gluon at one-loop level.
This remarkable result was obtained by the correct
infra-red regularization of the effective action
which respects the causality.
With this infra-red regularization by causality one has
the dual Meissner effect and the magnetic confinement
of color in $SU(2)$ QCD \cite{cho3}.

The new result sharply contradicts with the earlier results,
in particular on the stability of the monopole condensation.
This has resurrected the old controversy on the stability of monopole
condensation. Since the stability of the monopole condensation is
such an important issue for the confinement in QCD, it is 
imperative that one looks for an independent method to resolve 
the controversy.  

A remarkable point of
gauge theories is that in the massless limit
the imaginary part of the effective
action is proportional to $g^2$, where $g$ is the gauge coupling
constant. This has been shown to be true in both
QCD \cite{savv,niel,ditt,cho3} and massless QED \cite{cho01,cho5}.
This allows us
to check the presence (or the absence) of the imaginary part
by a perturbative method, and recently two of us have presented 
a straightfoward perturbative method to resolve the controversy 
on the stability of monopole condensation
once and for all \cite{cho4}. {\it The purpose of this 
paper is to discuss the perturbative method to calculate 
the imaginary part of the one-loop effective action 
in massles gauge theories in more detail. We test the validity of this method 
in massless QED, and apply this to $SU(2)$ QCD. 
We confirm that the QCD effective action 
indeed has no imaginary part
in the presence of the monopole background but has a negative
imaginary part in the presence of the color electric background}.
This endorses our result 
based on the infra-red regularization by causality \cite{cho3},
and guarantees the stability of the monopole condensation in QCD.

It should be pointed out that the first to dispute
the SNO effective action, and to suggest that
the imaginary part of the QCD effective action could be calculated
by a perturbative method, was Schanbacher \cite{sch}.
Unfortunately this remarkable suggestion has met with skepticism 
and thus been completely ignored, partly because this
work also lacked the desired gauge invariance but
more probably because this suggestion has never been
confirmed by a concrete example before.
In this paper we verify this suggestion in massless QED, 
and calculate the imaginary
part of the QCD effective action perturbatively
with two different methods.
The first method is to draw the relevant Feynman diagrams
and calculate the imaginary part at one-loop order. The second
is to apply Schwinger's method of the perturbative expansion \cite{schw}
of the QED effective action to QCD, and calculate the imaginary part
to the second order.
These methods reproduce the identical result, identical to 
what we obtained with the infra-red regularization by causality.

The paper is organized as follows. In Section II
we review the Abelian formalism of QCD to make
the similarity between QED and QCD more transparent. In Section III
we reproduce the one-loop effective action of QCD using
the Abelian formalism. In Section IV we compare our effective action
with the SNO effective action, and review the origin of the 
instability of the SNO vacuum. In Section V we calculate the imaginary
part of the effective action of massless QED perturbatively,
and demonstrate that the perturbative result produces an identical
result, identical to the non-perturbative result.
In Section VI we calculate the imaginary part of the QCD 
effective action perturbatively with the Feynman diagrams.
In Section VII (and in the Appendix) 
we apply Schwinger's method to QCD to reproduce 
the desired result. Finally in Section VIII we discuss
the physical implications of our findings.

\section{Abelian formalism of QCD} \label{sec:Cho1}

In this section we review the gauge-independent Abelianization
of QCD \cite{cho1,cho2}, which we need to calculate the imaginary part of
the effective action using the Schwinger's method.
For simplicity we will concentrate on the SU(2) QCD in this paper.
We start from the gauge-independent decomposition 
of the gauge potential into the restricted potential
$\hat A_\mu$ and the valence potential $\vec X_\mu$ \cite{cho1,cho2}, 
which has recently
been referred to as the Cho decomposition 
or Cho-Faddeev-Niemi-Shabanov decomposition \cite{fadd,shab}.
Let $\n$ be the unit
isovector which selects the color charge direction everywhere
in space-time, and let \cite{cho1,cho2}
\bea \label{eq:decompose}
& \vec{A}_\mu =A_\mu \n - \oneg \n\times\pro_\mu\n+\X_\mu\nonumber
         = \hat A_\mu + \X_\mu, \nn\\
& (A_\mu = \n\cdot \vec A_\mu,~ \n^2 =1,~ \hat{n}\cdot\vec{X}_\mu=0),
\eea
where $A_\mu$ is the ``electric'' potential.
Notice that the restricted potential $\hat A_\mu$ is precisely
the connection which leaves $\n$ invariant under parallel transport,
\bea
\D_\mu \n = \pro_\mu \n + g {\hat A}_\mu \times \n = 0.
\eea
Under the infinitesimal gauge transformation
\bea
\delta \n = - \vec \alpha \times \n  \,,\,\,\,\,
\delta \A_\mu = \oneg  D_\mu \vec \alpha,
\eea
one has
\bea
&&\delta A_\mu = \oneg \n \cdot \pro_\mu \valpha,\,\,\,\
\delta \hat A_\mu = \oneg \D_\mu \valpha  ,  \nn \\
&&\hspace{1.2cm}\delta \X_\mu = - \valpha \times \X_\mu  .
\eea
This tells that $\hat A_\mu$ by itself describes an $SU(2)$
connection which enjoys the full $SU(2)$ gauge degrees of
freedom. Furthermore the valence potential $\vec X_\mu$ forms a
gauge covariant vector field under the gauge transformation.
{\it But what is really remarkable is that the decomposition is
gauge independent. Once the gauge covariant topological field
$\hat n$ is chosen, the decomposition follows automatically,
regardless of the choice of gauge} \cite{cho1,cho2}.

Remember that $\hat{A}_\mu$ retains all the essential
topological characteristics of the original non-Abelian potential.
Clearly $\hat{n}$ defines $\pi_2(S^2)$
which describes the non-Abelian monopoles \cite{wu,cho80}, and
characterizes
the Hopf invariant $\pi_3(S^2)\simeq\pi_3(S^3)$ which describes
the topologically distinct vacua \cite{bpst,cho79}.
Furthermore $\hat{A}_\mu$ has a dual
structure,
\begin{eqnarray}
& \hat{F}_{\mu\nu} = \partial_\mu \hat A_\nu-\partial_\nu \hat A_\mu
+ g \hat A_\mu \times \hat A_\nu = (F_{\mu\nu}+ H_{\mu\nu})\hat{n}\mbox{,}
\nonumber \\
& F_{\mu\nu} = \partial_\mu A_{\nu}-\partial_{\nu}A_\mu \mbox{,}
\nonumber \\
& H_{\mu\nu} = -\dfrac{1}{g} \hat{n}\cdot(\partial_\mu
\hat{n}\times\partial_\nu\hat{n})
= \partial_\mu \tilde C_\nu-\partial_\nu \tilde C_\mu,
\end{eqnarray}
where $\tilde C_\mu$ is the ``magnetic'' potential of the monopoles
(Notice that one can always introduce the magnetic
potential since $H_{\mu \nu}$ forms a closed two-form
locally sectionwise) \cite{cho1,cho2}.
Thus, one can identify the non-Abelian monopole potential by
\bea \label{eq:vecC}
\vec C_\mu= -\dfrac{1}{g}\hat n \times \partial_\mu\hat n ,
\eea
in terms of which the magnetic field is expressed by
\bea
\vec H_{\mu\nu}=\partial_\mu \vec C_\nu-\partial_\nu \vec C_\mu+ g
\vec C_\mu \times \vec C_\nu = H_{\mu\nu}\hat n.
\eea
This provides the gauge independent separation of the monopole
field $H_{\mu\nu}$ from the color electromagnetic field $ F_{\mu\nu}$.
The monopole potential (6) has been referred to as 
the Cho connection \cite{fadd,shab}.

With the decomposition (\ref{eq:decompose}), one has
\bea
\vec{F}_{\mu\nu}&=&\hat F_{\mu \nu} + \D _\mu \X_\nu -
\D_\nu \X_\mu + g\X_\mu \times \X_\nu,
\eea
so that the Yang-Mills Lagrangian is expressed as
\bea
&{\cal L} = -\dfrac{1}{4} \vec F^2_{\mu \nu } \nn\\
&=-\dfrac{1}{4}
{\hat F}_{\mu\nu}^2 -\dfrac{1}{4}(\D_\mu\X_\nu-\D_\nu\X_\mu)^2
\nn \\
&-\dfrac{g}{2} {\hat F}_{\mu\nu} \cdot (\X_\mu \times \X_\nu)
-\dfrac{g^2}{4} (\X_\mu \times \X_\nu)^2.
\eea
This shows that the Yang-Mills theory can be viewed as
a restricted gauge theory made of the restricted potential,
which has the valence gluons as its source \cite{cho1,cho2}.

An important advantage of the decomposition (1) is that it can actually
Abelianize (or more precisely ``dualize'') the non-Abelian
gauge theory \cite{cho1,cho2}. To see this let
$(\hat n_1,~\hat n_2,~\hat n)$
be a right-handed orthonormal basis and let
\begin{eqnarray}
&\vec{X}_\mu =X^1_\mu ~\hat{n}_1 + X^2_\mu ~\hat{n}_2\mbox{,} \nn\\
&(X^1_\mu = \hat {n}_1 \cdot \vec X_\mu,~~~X^2_\mu =
\hat {n}_2 \cdot \vec X_\mu)            \nonumber
\end{eqnarray}
and find
\begin{eqnarray}
&\hat{D}_\mu \vec{X}_\nu =\Big[\partial_\mu X^1_\nu-g
(A_\mu+ \tilde C_\mu)X^2_\nu \Big]\hat n_1 \nn\\
&+ \Big[\partial_\mu X^2_\nu+ g (A_\mu+ \tilde C_\mu)X^1_\nu \Big]\hat{n}_2.
\end{eqnarray}
So with
\bea
& B_\mu = A_\mu + \tilde C_\mu , \nn\\
&X_\mu = \dfrac{1}{\sqrt{2}} ( X^1_\mu + i X^2_\mu ),
\eea
one could express the Lagrangian explicitly in terms of the dual
potential $B_\mu$ and the complex vector field $X_\mu$,
\begin{eqnarray} \label{eq:Abelian}
&{\cal L}=-\dfrac{1}{4} G_{\mu\nu}^2
-\dfrac{1}{2}|\hat{D}_\mu{X}_\nu-\hat{D}_\nu{X}_\mu|^2
+ ig G_{\mu\nu} X_\mu^* X_\nu \nn\\
&-\dfrac{1}{2} g^2 \Big[(X_\mu^*X_\mu)^2-(X_\mu^*)^2 (X_\nu)^2 \Big] \nn\\
&= -\dfrac{1}{4}(G_{\mu\nu} + X_{\mu\nu})^2
-\dfrac{1}{2}|\hat{D}_\mu{X}_\nu-\hat{D}_\nu{X}_\mu|^2,
\end{eqnarray}
where
\bea
& G_{\mu\nu} = F_{\mu\nu} + H_{\mu\nu}, 
~~~\hat{D}_\mu{X}_\nu = (\partial_\mu + ig B_\mu) X_\nu, \nn\\
& X_{\mu\nu} = - i g ( X_\mu^* X_\nu - X_\nu^* X_\mu ). \nonumber
\eea
Clearly this describes an Abelian gauge theory coupled to
the charged vector field $X_\mu$.
But the important point here is that the Abelian potential
$B_\mu$ is given by the sum of the electric and magnetic potentials
$A_\mu+\tilde C_\mu$.
In this form the equations of motion of $SU(2)$ QCD is expressed by
\begin{eqnarray} \label{eq:AbelianEOM}
&\partial_\mu(G_{\mu\nu}+X_{\mu\nu}) = i g X^*_\mu
({\hat D}_\mu X_\nu -{\hat D}_\nu X_\mu ) \nn\\
&- i g X_\mu ({\hat D}_\mu X_\nu - {\hat D}_\nu X_\mu )^*, \nn\\
&\hat{D}_\mu(\hat{D}_\mu X_\nu- \hat{D}_\nu X_\mu)=ig X_\mu
(G_{\mu\nu} +X_{\mu\nu}).
\end{eqnarray}
This shows that one can indeed Abelianize the non-Abelian theory
with our decomposition. The remarkable change in this Abelian
formulation is that here the topological field $\hat n$ is
replaced by the magnetic potential $\tilde C_\mu$ \cite{cho1,cho2}.

An important feature of this Abelianization is that
it is gauge independent, because here we have never fixed
the gauge to obtain this Abelian formalism. So one might
ask how the non-Abelian gauge symmetry is realized in this Abelian
formalism. To discuss this let
\bea
&\vec \alpha = \alpha_1~\hn_1 + \alpha_2~\hn_2 + \theta~\hat n, \nn\\
&\alpha = \dfrac{1}{\sqrt 2} (\alpha_1 + i ~\alpha_2), \nn\\
&\vec C_\mu = - \dfrac {1}{g} \hn \times \partial_\mu \hn
= - C^1_\mu \hn_1 - C^2_\mu \hn_2, \nn\\
&C_\mu = \dfrac{1}{\sqrt 2} (C^1_\mu + i ~ C^2_\mu).
\eea
Then the Lagrangian (\ref{eq:Abelian}) is invariant not only under
the active gauge transformation (4) described by
\bea \label{eq:active}
&\delta A_\mu = \dfrac{1}{g} \partial_\mu \theta -
i (C_\mu^* \alpha - C_\mu \alpha^*),
~~~&\delta \tilde C_\mu = - \delta A_\mu, \nn\\
&\delta X_\mu = 0,
\eea
but also under the following passive gauge transformation
described by
\bea \label{eq:passive}
&\delta A_\mu = \dfrac{1}{g} \partial_\mu \theta -
i (X_\mu^* \alpha - X_\mu \alpha^*), ~~~&\delta \tilde C_\mu = 0, \nn\\
&\delta X_\mu = \oneg \hD_\mu \alpha - i \theta X_\mu.
\eea
Clearly this passive gauge transformation assures the desired
non-Abelian gauge symmetry for the Abelian formalism.
This tells that the Abelian theory not only retains
the original gauge symmetry, but actually has an enlarged (both the
active and passive) gauge symmetries.
{\it But we emphasize that this is not the ``naive'' Abelianization
of QCD which one obtains by fixing the gauge.
Our Abelianization is a gauge-independent Abelianization.
Besides, here the Abelian gauge
group is $U(1)_e \otimes U(1)_m$, so that
the theory becomes a dual gauge theory} \cite{cho1,cho2}. This is
evident from (15) and (16). This Abelianized QCD will become important
in the following.

\section{Monopole Condensation: A Review}\label{sec:Cho2}

To demonstrate the validity of the Abelian formalism
we now calculate the effective action of QCD
with the Abelian formalism, and reproduce the same effective
action that we obtained with the non-Abelian formalism.
So, using the background field method \cite{dewitt,pesk}, we first divide
the gluon field into two parts, the slow-varying
classical part $B_\mu$ and the fluctuating quantum
part $X_\mu$, and identify $B_\mu$
as the classical background. In this picture the active
gauge transformation (\ref{eq:active}) is viewed as the background gauge
transformation and the passive gauge transformation (16) is viewed
as the quantum gauge transformation.
Now, we fix the gauge of the quantum gauge transformation by
imposing the following gauge condition to $X_\mu$,
\bea
&\hat D_\mu X_\mu =0, ~~~~~(\hat{D}_\mu X_\mu)^* =0 \nn\\
&{\cal L}_{gf} =- \dfrac{1}{\xi}
|{\hat D}_\mu X_\mu|^2.
\eea
Remarkably, under the gauge transformation (\ref{eq:passive}) 
the gauge condition
depends only on $\alpha$, so the corresponding
Faddeev-Popov determinant is given by
\be \label{eq:complexFP}
M_{FP} = \left| \begin{array}{cc}
\dfrac{\delta (\hat{D}_\mu X_\mu)}{\delta \alpha} &
\dfrac{\delta (\hat{D}_\mu X_\mu)}{\delta \alpha^*} \\
\dfrac{\delta (\hat{D}_\mu X_\mu)^*}{\delta \alpha} &
\dfrac{\delta (\hat{D}_\mu X_\mu)^*}{\delta \alpha^*}
\end{array} \right|.
\ee

With this gauge fixing
the effective action takes the following form,
\begin{widetext}
\bea \label{eq:effaction}
&\exp \Big[iS_{eff}(B_\mu) \Big] = \dfrac{}{} \int
{\cal D} X_\mu {\cal D} X_\mu^* {\cal D}c_1{\cal D}c_1^{\dagger}
{\cal D}c_2{\cal D}c_2^{\dagger} \exp \Big{\lbrace} \dfrac{}{} 
i\int \Big[ -\dfrac{1}{4}(G_{\mu\nu} + X_{\mu\nu})^2
-\dfrac{1}{2}|\hat{D}_\mu{X}_\nu-\hat{D}_\nu{X}_\mu|^2 \nn\\
&-\dfrac{1}{\xi} |\hat {D}_\mu X_\mu|^2
+ c_1^\dagger (\hat{D}^2
+ g^2X_\mu^* X_\mu )c_1 - g^2 c_1^\dagger
 X_\mu X_\mu c_2 + c_2^{\dagger}
(\hat{D}^2 + g^2X_\mu^* X_\mu )^*c_2
- g^2 c_2^{\dagger} X^*_\mu X^*_\mu c_1 ~\Big] d^4x \Big{\rbrace},
\eea
where $c_1$ and $c_2$ are the complex ghost fields.
To evaluate the integral
notice that the functional determinants of the valence gluon
and the ghost loops are expressed as
\bea
&{\rm Det}^{-\frac{1}{2}} K_{\mu \nu}\simeq
{\rm Det}[-g_{\mu \nu}
 (\hat D \hat D)+ 2ig G_{\mu \nu}],
~~~~~~~{\rm Det} M_{FP} = {\rm Det} [-(\hat D \hat D)]^2.
\eea
Using the relation
\bea
&G_{\mu \alpha} G_{\nu \beta} G_{\alpha \beta} = \dfrac{1}{2} G^2
G_{\mu \nu} +\dfrac{1}{2}(G \tilde G) {\tilde G}_{\mu \nu},
~~~~~~~({\tilde G}_{\mu \nu}=\dfrac{1}{2}{\epsilon}_{\mu\nu\rho\sigma}
G_{\rho\sigma}),
\eea
one can simplify the functional determinants of the gluon
and the ghost loops as follows,
\bea \label{eq:functDet}
& \ln {\rm Det}^{-\frac 1 2} K = \ln {\rm Det} [(-\hD^2+2a)(-\hD^2-2a)]
+ \ln {\rm Det} [(-\hD^2-2ib)(-\hD^2+2ib)],\nn\\
& \ln {\rm Det}M_{FP} = 2\ln {\rm Det}(-\hD^2),
\eea
where
\bea
a = \dfrac{g}{2} \sqrt {\sqrt {G^4 + (G \tilde G)^2} + G^2},
~~~~~~~b = \dfrac{g}{2} \sqrt {\sqrt {G^4 + (G \tilde G)^2} - G^2}. \nn
\eea
With this one has \cite{cho3,cho4}
\bea
\Delta S = i \ln {\rm Det} [(-\hD^2+2a)(-\hD^2-2a)]
+ i \ln {\rm Det} [(-\hD^2-2ib)(-\hD^2+2ib)] - 2i \ln {\rm Det}(-\hD^2).
\eea
One can evaluate the functional determinants and find
\bea \label{eq:effaction2}
&\Delta {\cal L} =  \dfrac{1}{16 \pi^2}  \int_{0}^{\infty}
\dfrac{dt}{t^{3-\epsilon}} \dfrac{ a b t^2 / \mu^4}{\sinh (at/\mu^2)
\sin (bt/\mu^2)} \Big[ \exp(-2at/\mu^2)+\exp(2at/\mu^2) \nn\\
&+\exp(2ibt/\mu^2)+\exp(-2ibt/\mu^2)-2 \Big],
\eea
\end{widetext}
where $\mu$ is a dimensional parameter.
This is exactly the same expression of the effective action
that we have obtained before, with the non-Abelian
formalism of QCD \cite{savv,ditt,cho3,cho4}.
This demonstrates that our Abelian formalism
is identical to the conventional non-Abelian formalism of QCD.

\begin{figure*}
\includegraphics{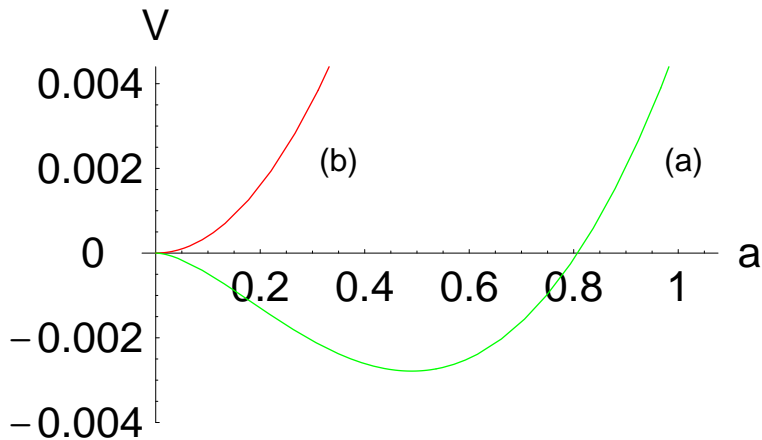}
\caption{\label{Fig. 1} The effective potential of SU(2) QCD in
the pure magnetic background. Here (a) is the effective potential
and (b) is the classical potential.}
\end{figure*}
\vspace{1cm}

The evaluation of the integral (24) for arbitrary 
$a$ and $b$ has been notoriously difficult \cite{savv,niel,ditt}.
Even in the case of ``simpler'' QED the integration of the
effective action has been completed only recently \cite{cho01,cho5}.
Fortunately the integral for the pure electric ($a=0$) and
pure magnetic ($b=0$) background has now 
been correctly performed \cite{cho3,cho4}.
For a pure monopole background (i.e., with $G_{\mu\nu}
=H_{\mu\nu}$) we have $b=0$,
so that the integral (24) becomes
\begin{widetext}
\bea
&\Delta{\cal L} = \dfrac{1}{16 \pi^2}\int_{0}^{\infty}
\dfrac{dt}{t^2} \dfrac{a/ \mu^2}{\sinh (at/\mu^2)}
\Big[\exp (-2at/\mu^2 )+  \exp (+2at/\mu^2) \Big].
\eea
To perform the integral we have to regularize the infra-red
divergence first. For this we go to the Minkowski time with
the Wick rotation, and find \cite{cho3,cho4}
\bea \label{eq:deltaL}
&\Delta {\cal L}=  \Delta{\cal L_+} + \Delta{\cal L_-}, \nn\\
& \Delta{\cal L_+} =  - \dfrac{1}{16 \pi^2}\int_{0}^{\infty}
\dfrac{dt}{t^2} \dfrac{a /\mu^2}{\sin (at/\mu^2)}
\exp (-2i a t/\mu^2 ), \nn\\
& \Delta{\cal L_-} =  - \dfrac{1}{16 \pi^2}\int_{0}^{\infty}
\dfrac{dt}{t^2} \dfrac{a/\mu^2}{\sin (at/\mu^2)}
\exp (+2i a t/\mu^2 ).
\eea
\end{widetext}
In this form the infra-red divergence has disappeared,
but now we face an ambiguity in choosing the correct contours
of the integrals in (\ref{eq:deltaL}). Fortunately this ambiguity can
be resolved by causality. To see this notice that the two integrals
$\Delta{\cal L_+}$ and $\Delta{\cal L_-}$ originate from the
two determinants ${\rm Det} (-\hD^2+2a)$ and ${\rm Det} (-\hD^2-2a)$,
and the standard causality argument requires us to
identify $2 a$ in the first determinant as
$2 a -i\epsilon$ but in the second determinant as
$2 a +i\epsilon$. {\it This tells that
the poles in the first integral in (\ref{eq:deltaL}) should lie above
the real axis, but the poles in the second integral should lie
below the real axis. From this we conclude
that the contour in $\Delta{\cal L_+}$ should pass below the
real axis, but the contour in $\Delta{\cal L_-}$ should pass above the
real axis}. With this causality requirement the two integrals
become complex conjugate to each other. This guarantees that
$\Delta{\cal L}$ is explicitly real, without any imaginary part.

With this infra-red regularization by causality we obtain
\bea \label{eq:effLb0}
&{\cal L}_{eff} = - \dfrac{a^2}{2g^2} -\dfrac{11}{48\pi^2}a^2(\ln
\dfrac{a}{\mu^2}-c ), \nn\\
&c=1-\ln 2 -\dfrac {24}{11} \zeta'(-1, \frac{3}{2})=0.94556... ,
\eea
for a pure monopole background,
where $\zeta(x,y)$ is the generalized Hurwitz
zeta function \cite{cho3,cho4}.

For the pure electric background (i.e., with $a=0$)
the infra-red regularization by causality gives us \cite{cho3,cho4}
\bea \label{eq:effLa0}
{\cal L}_{eff} = \dfrac{b^2}{2g^2} +\dfrac{11b^2}{48\pi^2}
(\ln \dfrac{b}{\mu^2}-c)-i\dfrac{11b^2}{96\pi}.
\eea
Observe that (\ref{eq:effLb0}) and (\ref{eq:effLa0}) are
related by the duality. In fact we can obtain one from the other
simply by replacing $a$ with $-ib$ and $b$ with $ia$.
This duality, which states that the effective action should be
invariant under the replacement
\bea
a \rightarrow - ib,~~~~~~~b \rightarrow ia,
\eea
was recently established as a
fundamental symmetry of the effective action of gauge theory,
both Abelian and non-Abelian \cite{cho3,cho01}. We emphasize that
the duality provides a very useful tool to check the self-consistency
of the effective action.

Remarkably the the effective action (27) generates the much desired
dimensional transmutation in QCD, the phenomenon 
Coleman and Weinberg first observed in massless scalar QED \cite{cole}.
It is this dimensional transmutation that produces the dynamical
generation of mass gap through the monopole condensation \cite{cho3,cho4}.
To demonstrate this we first renormalize the effective
action. Notice that the effective action (27) 
provides the following effective potential
\bea
V=\dfrac12 \dfrac{a^2}{g^2}
\Big[1+\dfrac{11 g^2}{24 \pi^2}(\ln\dfrac{a}{\mu^2}-c)\Big].
\eea
So we define the running coupling $\bar g$ by \cite{savv,cho3}
\bea
\frac{\partial^2V}{\partial a^2}\Big|_{a=\bar \mu^2}
=\frac{1}{ \bar g^2}.
\eea
With the definition we find
\bea
\frac{1}{\bar g^2} =
\frac{1}{g^2}+\frac{11}{24 \pi^2}( \ln\frac{{\bar\mu}^2}{\mu^2}
- c + \dfrac{3}{2}),
\eea
from which we obtain the following $\beta$-function,
\bea
\beta(\bar\mu)= \bar\mu \dfrac{\partial \bar g}{\partial \bar\mu}
= -\frac{11}{24\pi^2} \bar g^3~.
\eea
This is exactly the same $\beta$-function that one obtained
from the perturbative QCD to prove the asymptotic freedom
\cite{wil}. This confirms that our effective action is
consistent with the asymptotic freedom.

The fact that the $\beta$-function obtained from
the effective action becomes identical to the one obtained by
the perturbative calculation is really remarkable, because
this is not always the case. In fact in QED it has been
demonstrated that the running coupling and the $\beta$-function
obtained from the effective action is different from those
obtained from the perturbative method \cite{cho3,cho5}.

In terms of the running coupling the renormalized potential is given by
\bea
V_{\rm ren}=\dfrac{1}{2} \dfrac{a^2}{\bar g^2} 
\Big[1+\dfrac{11}{24 \pi^2 } \bar g^2
(\ln\dfrac{a}{\bar\mu^2}-\dfrac{3}{2})\Big],
\eea
which generates a non-trivial local minimum at
\bea
<a>=\bar \mu^2 \exp\Big(-\frac{24\pi^2}{11\bar g^2}+ 1\Big).
\eea
Notice that with ${\bar \alpha}_s = 1$ we have
\bea
\dfrac{<a>}{{\bar \mu}^2} = 0.48988... .
\eea
{\it This is nothing but the desired magnetic condensation.
This proves that the one
loop effective action of QCD in the presence of the constant magnetic
background does generate a phase transition thorugh the
monopole condensation} \cite{cho3,cho4}.

The corresponding effective potential is plotted in Fig. 1,
where we have assumed $\bar \alpha_s = 1, ~\bar \mu =1$.
The effective potential clearly shows that there is indeed a dynamical
generation of mass gap which indicates the existence of 
the confinement phase in QCD.

\section{Stability of Monopole Condensation}

The effective action of QCD in the presence of
pure magnetic background (i.e., for $b=0$) has first been calculated
by Savvidy and subsequently by Nielsen and Olesen \cite{savv,niel}.
Their effective action was almost identical to ours, 
except that theirs contains 
the following extra imaginary part,
\bea
{\rm Im} \thinspace {\cal L}_{eff} \Big |_{SNO} =
~\dfrac {a^2} {8\pi} ~~~~~~~b=0.
\eea
This sharply contradicts with our result (27) and (28),
which has the following imaginary part,
\bea \label{eq:imagine}
{\rm Im} ~{\cal L}_{eff}=\left\{{~~~~0~~~~~~~~~~~~~~~ b=0~~
\atop -\dfrac{11b^2}{96\pi}~~~~~~~~~~~a=0~.}\right.
\eea
Clearly the imaginary part of the SNO effective action
destabilizes our vacuum (35) through the pair creation of
gluons. This would nullify our result,
the monopole condensation and the dynamical generation of mass
in QCD. So it becomes a most urgent issue to find out
which of the effective actions is correct.

To settle thiss issue, one must understand the origin 
of the difference. The argument for the imaginary part in
the SNO effective action was that
the magnetic background should generate unstable tachyonic modes
in the long distance region. These tachyonic modes, they argued,
would generate an imaginary part to the effective action and
render the vacuum condensation unstable. And indeed one does
obtain (37), if one naively regularizes the infra-red divergence
of the integral (25) with the $\zeta$-function 
regularization \cite{niel,ditt}.

Notice that these unstable modes, however, are unphysical 
which violate the causality. Furthermore, the SNO vacuum
is not gauge invariant, and one can argue that
the unstable modes originate from this fact. 
So one must be careful to exclude the unphysical modes
when one regularizes the infra-red divergence.
As we have pointed out, the infra-red regularization
by causality naturally removes the unphysical modes and gives
us the correct effective action. Furthermore, this regularization
has been shown to preserve the duality, an important
consistency condition for the effective actions in
gauge theories \cite{cho3,cho4}.

Our infra-red regularization by causality 
assures the existence of the monopole
condensation in $SU(2)$ QCD without doubt. On the other hand, since
the $\zeta$-function regularization is also a well-established
regularization method, it is not at all clear which
regularization is the correct one for QCD at this moment.
Given the fact that the stability of the vacuum 
condensation is such an important
issue which can make or break the confinement, 
one must find an independent method 
to resolve the controversy.
We now discuss a straightforward perturbative method to calculate
the imaginary part of the effective action which can
resolve the controversy and
guarantee the stability of the monopole condensation. 
We present two independent
methods based on the perturbative expansion, the Feynman diagram
calculation and the perturbative calculation of 
the imaginary part of the effective action, to justify
the infra-red regularization by causality.

\section{Imaginary Part of Effective Action in Massless QED}

Before we discuss QCD we have to prove that one can indeed calculate
the imaginary part of the effective action either perturbatively
or non-perturbatively in massless gauge theories.
As far as we understand, this has never been demonstrated before.
So in this section we first calculate the imaginary part of the
effective action of massless QED perturbatively, and show
that the perturbative result is identical to the non-perturbative
result.

The imaginary part of the non-perturbative one-loop effective action
of QED has been known to be \cite{cho01,cho5},
\bea
&{\rm Im}\, {\cal L}_{QED} = \dfrac {ab}{8 \pi^2} \sum_{n=1}^{\infty}
\dfrac{1}{n} \coth \Big (\dfrac{n \pi a}{b}\Big ) \nn\\
&\times \exp \Big (-\dfrac {n \pi m^2}{b}\Big ),
\eea
where $m$ is the electron mass and
\bea
a = \dfrac{e}{2} \sqrt {\sqrt {F^4 + (F \tilde F)^2} + F^2}, \nn \\
b = \dfrac{e}{2} \sqrt {\sqrt {F^4 + (F \tilde F)^2} - F^2}. \nn
\eea
In the massless limit
this gives us \cite{cho01,cho5}
\bea
{\rm Im} \,{\cal L}_{QED} {\Big |}_{m=0} = \left\{{~0~~~~~~~~~~~~~~ b=0~~
\atop \dfrac {b^2}{48 \pi}~~~~~~~~~~~a=0~.}\right.
\eea
Notice that the imaginary part is indeed of the order $e^2$,
which is not the case when $m \ne 0$.
We emphasize that the massless limit is well-defined
because the one-loop effective action of massless QED has
no infra-red divergence when $ab=0$ \cite{cho01,cho5}.

\begin{figure*}
\includegraphics{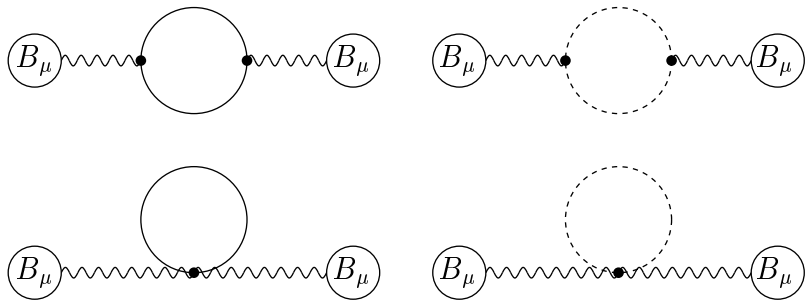}
\caption{The Feynman diagrams that contribute to the effective
action at $g^2$ order.\label{fig:Abelian}}
\end{figure*}
\vspace{1cm}

Now we calculate the imaginary part perturbatively to the order
$e^2$. There are two different ways to do this, by calculating the
Feynman diagram directly to the order $e^2$ or by making the
perturbative expansion of the non-perturbative one-loop effective
action. We start with the Feynman diagram. It is
well-known that, to the order $e^2$, the electron loop
diagram for an arbitrary background $A_\mu$
(with the dimensional regularization) gives us \cite{pesk}
\bea
&\dfrac{1}{2} \int d^4p A_\mu (p) \Pi_{\mu \nu}(p) A_\nu (-p) \nn \\
&= \dfrac{i e^2}{4\pi^2} \int d^4p A_\mu (p) (p^2 g_{\mu\nu} - p_\mu p_\nu)
A_\nu (-p) \nn \\
&\times \dfrac{}{} \int_0^1  dx  x(1-x) \Big (-\dfrac{2}{\epsilon} +
\gamma +
\ln [1 - x (1-x) \dfrac{p^2}{m^2}]\Big ) \nn\\
&=\dfrac{i e^2}{8 \pi^2} \int d^4p
F_{\mu\nu}(p)F_{\mu\nu}(-p) \dfrac{}{} \int_0^1 dx x(1-x) \nn\\
& \times \Big (-\dfrac{2}{\epsilon} + \gamma
+ \ln [1 - x (1-x) \dfrac{p^2}{m^2}]\Big ),
\eea
where $\Pi_{\mu\nu}(p)$ is the vacuum polarization tensor and
$m$ is the electron mass.
In the massless limit (with the subtraction of infra-red
divergence at $\mu^2$) this gives us
the leading contribution to the effective action
\bea
&\Delta S_{QED}=\dfrac{e^2}{48 \pi^2} \int d^4p
F_{\mu\nu}(p)F_{\mu\nu}(-p) \nn\\
&\times \Big [ \ln \Big ( \dfrac {p^2}{\mu^2} \Big )+ C_0
\Big ]
\eea
where $C_0$ is a regularization-dependent constant
which can be absorbed to the wave function renormalization.

Clearly the imaginary part could only
arise from the term $\mbox{ln} (p^2/\mu^2)$, so that
for a space-like $p^2$ (with $\mu^2>0$) the effective action
has no imaginary part. But since a space-like
$p^2$ corresponds to a magnetic background ($b=0$), we find that
the magnetic condensation generates no imaginary part,
at least at the order $e^2$.

As for the electric background ($a=0$) in general
we need to consider a time-like $p^2$, but here we are interested
in a constant electric background. In this case we have to go
to the limit $p^2=0$
and make the analytic continuation
of (42) to $p^2=0$. This is a delicate task. Fortunately
the causality, with the familiar prescription
$p^2 \rightarrow p^2-i \epsilon$, guides us to make
the analytic continuation correctly. Indeed the causality tells that
\bea
&\dfrac{}{}\lim_{p^2 \to0} ~\mbox{ln} \Big(\dfrac{p^2}{\mu^2}\Big) \rightarrow
\lim_{p^2 \to0} ~\mbox{ln} \Big(\dfrac{p^2}{\mu^2} -i\epsilon \Big) \nn\\
&=~\infty - i \dfrac{\pi}{2}.
\eea
From this we obtain (to the order $e^2$)
\bea
{\rm Im} ~{\cal L}_{QED} {\Big |}_{m=0} = \left\{{~0
~~~~~~~~~~~~~~ b=0,
\atop \dfrac{b^2}{48 \pi} ~~~~~~~~~~~a=0,}\right. \nn
\eea
which obviously is identical to the non-perturbative result (40).

Next, we calculate the imaginary
part of the effective action of massless QED
by making the perturbative expansion of the effective action
to the order $e^2$. For this it is instructive for us to review Schwinger's
perturbative calculation of the QED effective action,
who obtained \cite{schw}
\bea
&\Delta S_{QED}=\dfrac{e^2}{16 \pi^2} \int d^4p
F_{\mu\nu}(p)F_{\mu\nu}(-p) \nn\\
&\times \dfrac{}{}\int_{0}^{1} dv \dfrac{v^2 (1- v^2/3)}{1- v^2 + 4m^2/p^2}.
\eea
Clearly the imaginary part can only arise when the integrand 
develops a pole, so that the effective action has no imaginary part
when $p^2 > -4m^2$. This confirms that, for the magnetic background,
the QED effective action has no imaginary part.

But notice that
when $p^2<-4m^2$ the integrand develops a pole at
$v^2=1+4m^2/p^2$, and Schwinger 
taught us how to calculate the imaginary part
of the effective action with the causality prescription
$ p^2 \rightarrow p^2-i\epsilon$. But we notice that in the massless limit,
the pole moves to $v=1$. In this case the pole contribution
to the imaginary part is reduced by a half because only
one quarter, not a half, of the pole contributes to the
integral.  With this observation we again reproduce (40) \cite{cho4}.

This proves that one can calculate the imaginary part
of the effective action of massless QED perturbatively or
non-perturbatively, with identical results.
We emphasize that the perturbative calculation of the imaginary part
of the effective action of massless QED has never been
performed correctly before.

\section{Diagrammatic Calculation of Imaginary Part of QCD
Effective Action}
\label{sec:Feynman}
A remarkable point of QCD is that the imaginary part of the one-loop
effective action is of the order $g^2$. This is evident from
both (37) and (38). This allows us to calculate the imaginary part
perturbatively. In this section we calculate the imaginary part of the
effective action of QCD perturbatively from Feynman diagrams.
To do this we can either use the non-Abelian formalism
or the Abelian formalism of QCD. Here we will use
the latter in order to make the connection
between QED and QCD more transparent.

For an arbitrary background $B_\mu$ there are four Feynman
diagrams that contribute to the order $g^2$. They are shown
schematically in Fig.~\ref{fig:Abelian}.
Here the straight line and the dotted line represent the
valence gluon and the ghost, respectively.
The tadpole diagrams contain a quadratic
divergence which does not appear in the final result.

The sum of these diagrams
is given (in Feynman gauge with dimensional regularization)
by \cite{pesk}
\bea
&\dfrac{1}{2} \int d^4p B_\mu (p) \Pi_{\mu \nu}(p) B_\nu (-p) \nn \\
&=i \dfrac{11 g^2}{96 \pi^2} \int d^4p
G_{\mu\nu}(p)G_{\mu\nu}(-p) \dfrac{}{} \int_0^1 dx x(1-x) \nn\\
& \times \Big (-\dfrac{2}{\epsilon} + \gamma 
+ \ln [1 - x (1-x) \dfrac{p^2}{\mu^2}]\Big ),
\eea
so that we have
\bea \label{eq:Bcorrection}
&\Delta S_{eff} = - \dfrac{11g^2}{96\pi^2} \displaystyle \int d^4p
G_{\mu \nu}(p) G_{\mu \nu}(-p) \nn\\
&\times \left[\dfrac{2}{\epsilon} - \gamma - \mbox{ln}
\left(\dfrac{p^2}{\mu^2}\right) \right].
\eea
Evidently the imaginary part can only
arise from the term $\mbox{ln} (p^2/\mu^2)$, and just
as in massless QED we have no imaginary part for
the monopole condensation (for $b=0$).

To evaluate the imaginary part
for a constant electric background (for $a=0$) 
we have to make the analytic continuation
of (46) to $p^2=0$. Here again the causality requirement (43) dictates 
how to make the analytic continuation, and we obtain
\bea
{\rm Im} ~\Delta {\cal L} = -\dfrac{11 b^2}{96 \pi} ~~~~~~~~~(a=0).
\eea
So (46) gives us \cite{cho4}
\bea 
{\rm Im} ~{\cal L}_{eff}=\left\{{~~~~0~~~~~~~~~~~~~~~ b=0~~
\atop -\dfrac{11b^2}{96\pi}~~~~~~~~~~~a=0~,}\right. \nn
\eea
which is identical to (38). From this 
we conclude that indeed the Feynman diagram calculation
of the imaginary part does produce the result which is
identical to what we obtained from the one-loop effective 
action of QCD \cite{cho3,cho4}.

\section{Schwinger's Method} \label{sec:Schwinger}

To remove any remaining doubt about the imaginary part (\ref{eq:imagine})
we now verify
the above result by evaluating the functional determinant
(23) of the effective action
perturbatively to the order $g^2$. With our Abelian
formalism of QCD, one can do this by following
the method Schwinger used to evaluate
the effective action of QED perturbatively step by step \cite{schw}. 
This is presented in
the Appendix. Here we do this in a slightly different way
using a modern language, by calculating the functional
determinant (23) to the order $g^2$.

To calculate the determinant we consider the determinant of the form
\bea
{\rm Det} \,M &=& {\rm Det}\, (-\hD^2 +g R),
\eea
where $R = \pm 2a/g, \pm 2ib/g$, and $0$.
We can rewrite
\bea
&{\rm Det} \, M = {\rm Det} \, (-\partial^2) \cdot {\rm Det}
[1 + \dfrac{1}{\partial^2} ig ({\partial}_\mu B_\mu + B_\mu \partial_\mu) \nn\\
&- \dfrac{1}{\partial^2} g^2 B_\mu^2 - \dfrac{1}{\partial^2} g R]  \nn \\
&= {\rm Det} \, (-\partial^2) \cdot {\rm Det} [1 + \Delta + \tilde \Delta
+\hat \Delta],
\eea
and calculate the determinant by expanding it.
At the lowest order in $g^2$ we have
\bea
\ln {\rm Det} M ={\rm Tr} \ln M = {\rm Tr} (\tD
- \dfrac{1}{2} \Delta^2 - \dfrac{1}{2} \hat \Delta^2).
\eea
Now, using a standard technique \cite{pesk,hon}
(with dimensional regularization)
we obtain to the order $g^2$,
\bea
& {\rm Tr} \ln\, (-\hD^2 +g R) \nn \\
&  = \dfrac{i g^2}{ 16 \pi^2} \int d^4 p   \int_0^1 dv
\int_0^\infty \dfrac{dt}{t} \exp \Big [{-\dfrac{p^2}{4 \mu^2} (1 - v^2) t}
     \Big ] \nn \\
& \times \Big [\dfrac{v^2}{4} G_{\mu \nu}(p) G_{\mu \nu }(-p)
 - \dfrac{1}{2} R(p) R(-p) \Big ].
\eea
From this we find that the determinant (23) is given,
up to the order $g^2$, by \cite{cho4}
\bea\label{eq:final1}
&\Delta S = \dfrac{g^2}{8\pi^2}
\int d^4p G_{\mu \nu}(p) G_{\mu \nu}(-p) \Sigma (p), \nn\\
&\Sigma (p) = \dfrac{}{}\int_{0}^{1} dv (1-\dfrac{v^2}{4})
\dfrac{}{}\int_{0}^{\infty}\dfrac{dt}{t}
\exp[-\dfrac{p^2}{4 \mu^2}(1-v^2) t] \nn\\
&= -2\dfrac{}{}\int_{0}^{1} dv \dfrac{v^2(1-v^2/12)}{1-v^2} + C_1,
\eea
where $C_1$ is a regularization-dependent constant.
Now, it is straightforward to evaluate the imaginary part of $\Delta S$,
because the imaginary part can only arise from the pole at $v=1$.
So repeating the argument to derive 
the imaginary part for massless QED in Section V,
we again reproduce (\ref{eq:imagine}), after the proper
wave function renormalization \cite{cho4}.

The fact that the Feynman diagram calculation and the Schwinger's
method produce an identical result is not accidental,
because they are physically equivalent \cite{pesk}.
In fact, one can easily convince oneself that, order by order 
the perturbative expansion of the determinant (23) has one to
one correspondence with the Feynman diagram expansion.
And up to the order $g^2$,
the Feynman diagrams that contribute to the determinant (23) are
precisely those given by Fig. 2. Another way to see this
is to notice that, from the definition of the exponential
integral function \cite{table}
\bea
&Ei (-z) = - \dfrac{}{} \int_{z}^{\infty} \dfrac{d\tau}{\tau} \exp (-\tau) \nn\\
&= \gamma + \mbox{ln} z + \dfrac{}{} \int_{0}^{z} \dfrac{d\tau}{\tau}
\left[\exp (-\tau)-1\right],~~({\rm Re} z > 0)
\eea
we can express $\Sigma (p)$ in (52) by
\bea
&\Sigma (p) = - \dfrac{}{}\lim_{\epsilon\to0} \int_{0}^{1} dv
(1-\dfrac{v^2}{4}) \left[ \gamma + \mbox{ln} \Big(\dfrac{p^2}{4 \mu^2}(1-v^2)
\epsilon \Big) \right] \nn\\
&=-\dfrac{11}{12}\left[\mbox{ln}\left(\dfrac{p^2}{\mu^2}\right)
+ C_2 \right],
\eea
where $C_2$ is another constant
which can be removed by the wave function renormalization.
But this is exactly what we had in (46) in the Feynmam diagram 
calculation (up to the irrelevant renormalization constant).
This assures that the calculations of the imaginary part
by Feynman diagram and by Schwinger's
method are indeed physically equivalent.

It must be emphasized again that, just as in the calculation of 
the non-perturbative effective action, 
the causality plays the crucial role
in these perturbative calculations of
the imaginary part of the effective action.

{\it This confirms that the monopole condensation
indeed describes a stable vacuum, but the electric background creates
the pair-annihilation of the valence gluons in $SU(2)$ QCD} \cite{cho3,cho4}.
As importantly this confirms that the infra-red regularization
by causality in the calculation of the effective action of QCD
is indeed correct.

\section{Discussion}

The proof of the monopole condensation in QCD
has been extremely difficult to attain. The earlier attempts
to calculate the effective action of QCD to prove the confinement
have produced a negative result.
In particular, the instability of the SNO vacuum has created a
false impression that one can not prove the monopole condensation
by calculating the effective action of QCD.
The instability of the SNO vacuum can be traced back to 
the $\zeta$-function regularization of the infra-red divergence
of the gluon loop integral (25). But we have recently 
pointed out that the $\zeta$-function regularization violates
the causality, and proposed the infrared regularization
by causality as the correct regularization in QCD.
With the infrared regularization by causality we 
proved that a stable monopole condensation does take place
at one loop level which can generate the dual Meissner effect and
the confinement of color in $SU(2)$ QCD \cite{cho3,cho4}. 

Considering the fact that the $\zeta$-function regularization 
is a well-established regularization method which has worked
so well in other theories, however, one can not easily dismiss 
the instability of the SNO vacuum. Under this circumstance
one definitely want to know which is the correct regularization
in QCD, and why that is so.

In this paper we have checked the correctness of 
the infrared regularization by causality with
two independent methods based on the perturbative expansion.
Both methods produce a result
identical to the one obtained with the infra-red regularization
by causality, confirming the stability of the monopole condensation.
This confirms that the magnetic
confinement is indeed the correct confinement mechanism in QCD.

We emphasize that the above perturbative calculation
of the  imaginary part of the one-loop effective action
was possible because in QCD (and in massless QED)
the imaginary part of the one-loop effective action is
of the order $g^2$ \cite{cho4,sch}. This assures us that one can make
a perturbative expansion for the imaginary part of the effective action.
For massive QED, for example, this calculation does not make sense because
the imaginary (as well as the real) part of the effective action
simply does {\it not} allow a convergent
perturbative expansion \cite{cho01,cho5}. The same argument
applies to the real part of QCD. Only
for the imaginary part of massless gauge theories one can make
sense out of the perturbative calculation.

One might worry about the negative signature of the imaginary part
in the QCD effective action. To understand the origin
of this, compare QCD with
QED. In QED we integrate the electron loop, and
we know that the imaginary part
of the effective action is positive \cite{cho01,schw}.
But in QCD we have the valence gluon loop, and 
obviously the electron and the valence gluon 
have opposite statistics. Furthermore one can easily convince oneself
that the two loop integrals do not change the signature,
due to the the similar nature of interactions.
This means that the imaginary part of the effective action
in QED and QCD should have opposite signature. This is the reason for
the negative signature \cite{cho3,cho4}. 
We emphasize that this is what one should have expected
from the asymptotic freedom. A positive 
imaginary part would make pair-creation, 
not pair-annhilation, of the valence gluons. This would
produce the screenig, not the anti-screening, of the color charge.
Obviously this is against the asymptotic freedom.
This assures us that indeed the electric background generates
pair-annihilation, not pair-creation, of gluons in QCD.

We have neglected the quarks in this paper. We simply remark
that the quarks, just like in asymptotic freedom \cite{wil},
tend to destabilize the monopole condensation. In fact the stability
puts exactly the same constraint on the number of quarks as
the asymptotic freedom \cite{cho6}.
Furthermore here we have considered only the pure magnetic
or pure electric background. So, to be precise,
the above result only proves the existence of a stable
monopole condensation for a pure magnetic background. To show that
this is the true vacuum of QCD, one must calculate the effective action
with an arbitrary background in the presence of the
quarks and show that the monopole condensation
remains a true minimum of the effective potential. Fortunately, one
can actually calculate the effective action with an arbitrary
constant background, and show that indeed the monopole condensation
becomes the true vacuum of $SU(2)$ QCD, at least at one-loop
level \cite{cho6}.

It is truly  remarkable (and surprising) that the principles of
quantum field theory allow us to demonstrate confinement
within the framework of QCD.
The failure to establish confinement
within the framework of QCD has recently encouraged an increasing
number of people to entertain the idea that
a supersymmetric generalization of QCD might be necessary
for confinement \cite {witt}. Our analysis shows that
this need not be the case. One can indeed demonstrate that
QCD generates confinement by itself, with the existing
principles of quantum field theory.
This should be interpreted as a
most spectacular triumph of quantum field theory itself.

{\bf Appendix}

With the Abelian formalism of QCD, we can follow Schwinger
step by step to obtain the imaginary part of
the QCD effective action. For the pure magnetic case (i.e., for $b=0$),
(23) is written as
\begin{equation} \label{eq:deltaS}
\Delta S = i \mbox{Tr} ~[\ln (-\hat{D}^2 + 2a)
+ \ln (-\hat{D}^2 - 2a)].
\end{equation}
Using the identity
\begin{equation} \label{eq:lnDet}
\ln  M = - \int_0^\infty \dfrac{ds}{s} \exp (-iMs),
\end{equation}
this can be expressed as
\begin{eqnarray} \label{eq:expandS}
\Delta S &=& -i \mbox{Tr} \int_0^\infty \dfrac{ds}{s}
[\exp(-i(\hat{D}^2 -2a)s) \nn\\
&&+ \exp(-i(\hat{D}^2 +2a)s)].
\eea
The perturbative expansion of
\begin{equation} \label{eq:perturb}
\mbox{Tr}~U(s)_{\pm} \equiv \mbox{Tr} \exp(-i(\hat{D}^2 \mp 2a)s),
\end{equation}
in powers of $g\tilde{C}_\mu$ and $a$ (remember that $a$ contains a
factor of $g$) can be read from Schwinger's work \cite{schw}.

To second order in $g$ we have
\begin{eqnarray} \label{eq:TrU}
&\mbox{Tr}~U(s)_{\pm} = \mbox{Tr}U_0(s)
-is\mbox{Tr}[{\mathcal G}_\pm U_0(s)] \nonumber \\
&  - \dfrac{1}{2} s^2
\int_0^1 du \mbox{Tr}[{\mathcal G}_\pm U_0((1-u)s)
{\mathcal G}_\pm U_0(us)],
\end{eqnarray}
where
\begin{eqnarray}
&U_0(s) = \exp (i\partial^2 s), \nn \\
&{\mathcal G}_\pm = - \hat{D}^2 + \partial^2 \pm 2a. \nn
\end{eqnarray}
Substituting these into (\ref{eq:expandS}) we have
\begin{widetext}
\begin{eqnarray}
&\Delta S =
-2i\displaystyle \int \dfrac{ds}{s} \Big{\lbrace} \mbox{Tr}
\Big[\exp (i\partial^2 s)\Big]
-is\mbox{Tr} \Big[g^2 \tilde{C}_\mu^2 \exp (i\partial^2 s)\Big] \nn\\
&+ g^2s^2 \displaystyle \int^1_{-1} dv \mbox{Tr}
\Big[\partial_\mu \tilde{C}_\mu
\exp \Big(\dfrac{i}{2}\partial^2(1-v)s\Big) \partial_\mu
\tilde{C}_\mu \exp \Big(\dfrac{i}{2}\partial^2(1+v)s\Big)\Big]
\nonumber \\
&- a^2s^2 \displaystyle \int^1_{-1} dv 
\mbox{Tr} \Big[\exp \Big(\dfrac{i}{2}\partial^2(1-v)s\Big)
\exp \Big(\dfrac{i}{2}\partial^2(1+v)s\Big)\Big] \Big{\rbrace},
\end{eqnarray}
where $u=(1+v)/2$.
The next step is to take the
trace, yielding
\begin{eqnarray}
&\Delta S = -2i\displaystyle
\int \dfrac{ds}{s}\int d^4p \int \dfrac{d^4k}{(2\pi)^4}
\Big{\lbrace} \mbox{Tr} \Big[ \exp (-ik^2 s)\Big]
-is\mbox{Tr} \Big[g^2 \tilde{C}_\mu^2 \exp (-ik^2 s)\Big] \nonumber \\
& -\displaystyle g^2s^2 \int^1_{-1} dv \mbox{Tr} \Big[k_\mu \tilde{C}_\mu
\exp \Big(-\dfrac{i}{2}\Big(k+\dfrac{p}{2}\Big)^2(1-v)s\Big) k_\mu
\tilde{C}_\mu \exp \Big(-\dfrac{i}{2}\Big(k-\dfrac{p}{2}\Big)^2(1+v)s
\Big)\Big] \nonumber \\
& - a^2s^2 \displaystyle \int^1_{-1} dv
\mbox{Tr} \Big[ \exp \Big(-\dfrac{i}{2}\Big(k+\dfrac{p}{2}\Big)^2(1-v)s
\Big)
\exp \Big(-\dfrac{i}{2}\Big(k-\dfrac{p}{2}\Big)^2(1+v)s\Big)\Big]
\Big{\rbrace}.
\end{eqnarray}
Following Schwinger's technique
and keeping only nonvanishing field-dependent terms, we obtain
\begin{eqnarray}
\Delta S = \dfrac{a^2}{4\pi^2}
\displaystyle \int \dfrac{ds}{s} \int d^4p 
\displaystyle \int_{0}^{1} dv \left(1 - \dfrac{v^2}{4} \right)
\exp \left(-\dfrac{i}{4} p^2(1-v^2)s\right).
\end{eqnarray}
Now, with the Wick rotation $s\rightarrow -is$ and the integration
by parts with respect to $v$, we have
\bea 
\Delta S = \dfrac{a^2}{8\pi^2} \int d^4p \int \dfrac{ds}{s}
\Big{\{} \dfrac{11}{6} + sp^2 \displaystyle \int_{0}^{1} dv v^2
\Big(1-\dfrac{v^2}{12} \Big) \exp
\Big(-\dfrac{p^2}{4}(1-v^2)s\Big)\Big{\}}  \nn 
\eea
\end{widetext}
The first term in the square bracket has no imaginary part and is
removed by renormalization. The second term is integrated over $s$
to yield 
\bea \label{eq:final} 
\Delta S = -\dfrac{a^2}{2\pi^2} \int d^4p 
\int_{0}^{1} dv \dfrac{v^2 (1-v^2/12 )}{1-v^2}. 
\eea

On the other hand for the pure electric background (i.e., for $a=0$),
we have
\begin{equation} \label{eq:final2}
\Delta S = i \mbox{Tr} ~[\ln (-\tilde{D}^2 + 2ib)
+ \ln (-\tilde{D}^2 - 2ib)],
\end{equation}
Repeating the steps that lead to (\ref{eq:final}) we find
\begin{equation} \label{eq:bfinal}
\Delta S = \dfrac{b^2}{2\pi^2}
\int d^4p \int_{0}^{1} dv
\dfrac{v^2 (1-v^2/12)}{1-v^2}.
\end{equation}
So, for an arbitrary background, we have
\bea
\Delta S = -\dfrac{a^2 -b^2}{2\pi^2} \int d^4p 
\int_{0}^{1} dv \dfrac{v^2 (1-v^2/12 )}{1-v^2}. \nn
\eea
This is identical to (52), again up to the irrelevant 
renormalization constant. 
This, with the argument in Sections VI and VII,
confirms that the Schwinger's method reproduces (38).

{\bf Acknowledgements}

~~~One of the authors (YMC) thanks S. Adler and F. Dyson
for the fruitful discussions, and Professor C. N. Yang for
the continuous encouragements. The other (DGP)
thanks Professor C. N. Yang for the fellowship at Asia Pacific
Center for Theoretical Physics, and appreciates Haewon Lee
for numerous discussions.
The work is supported in part by Korea Research Foundation (Grant KRF-2001
-015-BP0085) and by the BK21 project of Ministry of Education.

\end{document}